Gilbert Damping Parameters of Epitaxially-Stabilized Iron Gallium Thin Films from Ferromagnetic Resonance

Authors: Ruth Loh[1], Sujan Shrestha[1,2], Jiaxuan Wu[3], Jia-Mian Hu[3], Christoph Adelmann[2], Florin Ciubotaru[2], John T. Heron[1]

Affiliations:
[1]The Ferroelectronics Laboratory, University of Michigan, Ann Arbor, MI, USA
[2]Imec, Leuven, Belgium
[3]Department of Materials Science and Engineering, University of Wisconsin-Madison, Madison, WI, USA

Iron gallium (FeGa) alloys are excellent rare-earth-free magnetostrictors. Through epitaxial stabilization, the disordered A2 alloy can be extended from 19% to 30% gallium resulting in a magnetostrictive coefficient almost twice than that which is seen in rare earth magnetostrictors like $SmFe_2$. In a composite magnetoelectric structure, this makes epitaxially-stabilized iron gallium a key material for energy-efficient beyond CMOS technologies. The energy dissipation and speed of magnetoelectric switching, however, is affected by the magnetic resonance frequency and damping. Here we report the evolution of the ferromagnetic resonance and key materials parameters (magnetic anisotropy, magnetic damping, and magnetostriction coefficient) for 70 nm thick epitaxially-stabilized single crystal A2 FeGa films beyond 19% Ga. Using flip chip ferromagnetic resonance (1-14 GHz), we find that the Gilbert damping parameter spans the range of 0.09-0.16 and decreases as the Ga concentration increases. This correlates an increasing magnetoelastic coupling with a reduction in the Gilbert damping. We find that the effective damping is a mix of contributions from the intrinsic magnon-phonon scattering and other scattering/dissipation mechanisms, with the latter being dominant especially at high Ga composition. Our results provide insight into the mechanism of magnetic relaxation in metastable high magnetostriction materials and potential switching behavior of composite magnetoelectrics.

Prior work by Meisenheimer et al. demonstrated that the chemically disordered A2 phase of iron gallium ($Fe_{1-x}Ga_x$) can be epitaxially-stabilized to beyond 19 at% gallium and thereby extend the magnetostriction to 3500 ppm at 30 at%[1]. This is contrary to what happens in bulk, where higher percent gallium leads to an increase in the magnetostrictive coefficient until about x = 19 at%, where a phase change to a chemically ordered $D0_3$ FeGa phase suppresses the $B_1$ magnetoelastic coefficient and thus the magnetostriction[2-5]. In epitaxially-stabilized A2 thin films, the $B_1$ magnetoelastic coefficient continues to increase linearly from that at 19 at% all the way to x = 30 at%. However, the magnetostriction increases non-linearly as x approaches 30 at% as the shear modulus approaches 0. Building upon this, a large converse magnetoelectric effect (~$10^5$ s/m) was demonstrated when FeGa is epitaxially grown on PMN-PT single crystals. The large magnetoelectric coefficient suggests that significant energy efficiencies and fast switching times can be achieved in spin-based logic designs[6,7]. While these estimations have been made, it is not clear how the large magnetostriction in the material influences the Gilbert damping parameter of the $Fe_{1-x}Ga_x$ films. The Gilbert damping parameter has been measured in FeGa films where x~19 at%, but understanding of its evolution at higher Ga %, especially for epitaxially-stabilized A2 films, is lacking[8-11]. As the Gilbert damping parameter influences both the energy needed to switch and the speed of switching, it is necessary to know the experimental values of this parameter and understand how the various materials parameters interplay with it[12,13].

The Gilbert damping parameter is a significant parameter in the study of magnetization dynamics. These dynamics are governed by the Landau-Lifshitz-Gilbert (LLG) equation,

$$\frac{d\boldsymbol{M}}{dt} = -\gamma(\boldsymbol{M} \times \boldsymbol{H}_{eff}) + \frac{\alpha}{M_s}(\boldsymbol{M} \times \frac{d\boldsymbol{M}}{dt}) \qquad (1)$$

in which the change in a material's magnetization (***M***) is described by two torques: one that causes magnetization precession around a total effective magnetic field (***H**$_{eff}$*) and a damping (α) term that aligns the magnetization of the material with the total effective field. Within the first term is the gyromagnetic ratio γ, and within the damping (second) term is a phenomenological term known as the Gilbert damping parameter, α[14]. ***M**$_s$* is the saturation magnetization. Commonly, magnetization dynamics are measured by ferromagnetic resonance where empirical values represent an effective damping due to two contributions to the resonance width: intrinsic damping and two-magnon scattering[15]. Smaller values of α typically lead to longer precession times and lower energy consumption because less energy is lost to the lattice[16,17] or other excitation[18,19,20], while larger values of α will typically cause the magnetization to stabilize more quickly. As epitaxially-stabilized A2 FeGa films have enhanced magnetoelastic coupling and reduced differences in longitudinal and shear mechanical stiffness components as the Ga concentration is increased, it is important to understand how the damping parameter evolves and how energy-angular momentum are interconverted in magnetic relaxation. As both switching speed and energy dissipation are greatly influenced by α, it is an especially important parameter for spintronics-based beyond CMOS technology, where energy-delay performance is a critical figure of merit[12,21].

Here, we investigate 70 nm thick epitaxial Fe$_{1-x}$Ga$_x$ films in the chemically-disordered A2 phase deposited on MgO substrates with x varying from 19.5 at% to 30 at%. Using flip-chip ferromagnetic resonance (FMR) we determine the Gilbert damping parameter and other key materials parameters of these films such as the saturation magnetization, magnetic anisotropy energy, and magnetoelastic coupling coefficient. We find that the magnetization spans 713-1380 kA/m, measured magnetic anisotropy spans 0-31 kJ/m$^3$, and the Gilbert damping parameter spans 0.09-0.16 in this composition range with all parameters decreasing with increasing Ga fraction. The magnetostriction coefficient, however, increases with Ga fraction, spanning 23-320 ppm. These trends indicates that the Gilbert damping parameter negatively correlates with the magnetoelastic coupling coefficient and magnetostriction.

The Fe$_{1-x}$Ga$_x$ thin films were grown by pulsed laser deposition on (100)-oriented MgO substrates under vacuum at a temperature of 400 °C on top of a sub-nanometer thick seed layer of pure Fe grown at the same temperature. Following the deposition, the films were cooled to below 70 °C and capped with a thin layer of HfO$_2$. The X-ray diffraction 2θ-ω scan measurement (Figure 1(a)) confirms the presence of the well-defined 002 peak of the chemically-disordered A2 phase around 65° and the absence of other peaks (most notably the 100 peak from the D0$_3$ phase) reveals that the films are single crystalline and single phase. The plot in Figure 1(b) shows that the 2θ peak value shifts to lower values revealing an increasing out-of-plane lattice parameter as the Ga % increases. The out-of-plane lattice parameter for each composition (Figure 1(c)) follows a linear trend commensurate with Vegard's law and similar to what is reported in Ref. 22. Analysis of the intensity fringes observed in x-ray reflectivity measurements confirms that all samples are around 70 nm thick. Furthermore, the reciprocal space mapping was measured around the 022 reflection of the MgO substrate in which the $\bar{1}12$ peak of disordered A2 phase of FeGa is clearly observed revealing that the thin films are epitaxial and largely relaxed (Figure 1 (d)). Calculation of the lattice parameters reveals that the FeGa thin films are slightly tetragonal with an elongated c-axis and contracted a-axis. For example, the 27% Ga sample has a = 2.78 Å and c = 2.92 Å.

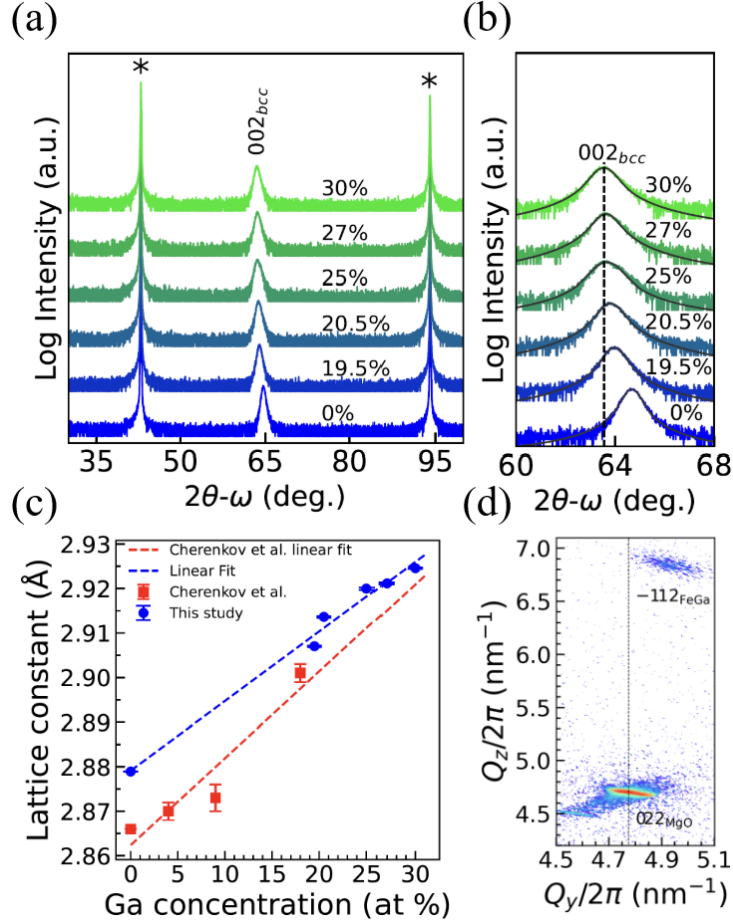

Figure 1: (a) X-ray diffraction 2θ-ω scan of epitaxially-stabilized FeGa films with atomic % Ga ranging from x = 0 to 30. The asterisk (*) symbol represents the 001 peaks for MgO substrates. (b) 2θ-ω scan about the 002 peaks of the FeGa films with different Ga concentration. Diffraction peaks shift to lower 2θ values as Ga % increases. (c) Out-of-plane lattice constants of FeGa samples compared to values found in literature[22]. The blue and red dashed lines represent the linear fit to data obtained in this study and Ref. 22, respectively, and are in relative agreement with expectations from Vegard's law. (d) Reciprocal space map of the 27 atomic % Ga sample measured around the 022 reflection of MgO substrate. The $\bar{1}12$ peak of the FeGa thin film reveals the relaxation. There is a slight tetragonality indicating a small residual strain.

FMR measurements were performed using a vector network analyzer (VNA) and vector electromagnet at room temperature from 1-14 GHz in a flip-chip configuration. The FeGa films were placed face down on a coplanar waveguide (CPW) and experienced fields up to 100 mT along the in-plane <100> easy axis orientation of FeGa (which is the <110> direction for the MgO substrate). The $HfO_2$ insulating capping layer prevented electrical contact between the sample and CPW. Using an rf power of -5 dBm, the $S_{21}$ parameter for each film was measured in order to estimate the damping parameter α. Before performing each measurement, a saturating field of 100 mT was applied in the plane of the sample and perpendicular to the CPW to serve as a background measurement. The background measurement was then subtracted from each of the measured $S_{21}$ curves. To extract physical parameters, a Kittel relationship was derived and fit to the data. We begin with equation (1) where the effective field $\boldsymbol{H}^{eff} = \boldsymbol{H}^{anis} + \boldsymbol{H}^{d} + \boldsymbol{H}^{mel} + \boldsymbol{H}^{bias}$ where $\boldsymbol{H}^{anis}$ is the magnetocrystalline anisotropy effective field, $\boldsymbol{H}^{d} = (0,0,-M_3)$ is the demagnetization field, $\boldsymbol{H}^{mel}$ is the magnetoelastic effective field, and $\boldsymbol{H}^{bias} = (H_1^{bias}, 0, 0)$ is the external bias field. As this model is studying the Kittel mode magnon (i.e. the wavevector **k** is zero), the exchange coupling field $\boldsymbol{H}^{exch}$ is not included. The subscripts, with '1' ∥ [100] ∥ x, '2' ∥ [010] ∥ y, '3' ∥ [001] ∥ z, indicate the orthogonal axes in the crystal physics coordinates system of the paramagnetic phase of FeGa and *x-y-z* indicates the lab coordinate system.

For a magnetic material with a cubic paramagnetic phase, $\boldsymbol{H}^{anis}$ is given by[23],

$$H_i^{anis} = -\frac{2m_i}{\mu_0 M_s}\left[K_1\left(m_j^2 + m_k^2\right)\right], \tag{2}$$

where high-order anisotropy coefficients ($K_2$ and above) are omitted and $i,j,k$=1,2,3. Here, the field is expressed in terms of components of the normalized magnetization vector $\boldsymbol{m} = \boldsymbol{M}/M_s = (m_1, m_2, m_3)$. $\boldsymbol{H}^{mel}$ is the magnetoelastic effective field given as[24],

$$H_i^{mel} = -\frac{2}{\mu_0 M_s}\left[B_1 m_i \varepsilon_{ii} + B_2\left(m_j \varepsilon_{ij} + m_k \varepsilon_{ik}\right)\right], \tag{3}$$

where the magnetoelastic coupling coefficients are $B_1 = -1.5\lambda_{100}(c_{11} - c_{12})$ and $B_2 = -3\lambda_{111}c_{44}$. Here $\lambda_{100}$ and $\lambda_{111}$ are the saturation magnetostriction along the <100> and <111> axes, respectively. $c_{11}$, $c_{12}$, and $c_{44}$ are the elastic stiffness coefficients (in Voigt notation) of the cubic FeGa. $\varepsilon_{ij}$ is obtained based on the thin-film boundary condition. Specifically, the in-plane components of a substrate-clamped epitaxial grown thin film are[25],

$$\varepsilon_{11} = \varepsilon_{11}^{res} = \frac{a^{measured} - a^{bulk}}{a^{bulk}}, \varepsilon_{22} = \varepsilon_{22}^{res} = \frac{b^{measured} - b^{bulk}}{b^{bulk}}, \varepsilon_{12} = 0, \tag{4}$$

where $a^{measured}$ is the in-plane lattice constant measured for the thin film sample, $a^{bulk}$ is the in-plane lattice constant for the higher-symmetry paramagnetic bulk crystal. Incorporating the stress-free boundary condition at the surface,

$$\sigma_{13} = \sigma_{23} = \sigma_{33} = 0, \tag{5}$$

where $\sigma_{ij} = c_{ijkl}(\varepsilon_{kl} - \varepsilon_{kl}^0)$ is the stress tensor, $\varepsilon_{kl}^0$ is the spontaneous strain arising from a paramagnetic-to-ferromagnetic phase transition, given by,

$$\varepsilon_{kk}^0 = \frac{3}{2}\lambda_{100}\left(m_k^2 - \frac{1}{3}\right), \varepsilon_{kl}^0 = \frac{3}{2}\lambda_{111}m_k m_l. \tag{6}$$

Combining Eqs. (5-6) allows us to obtain the remaining total strain component (also see Ref. 26),

$$\varepsilon_{13} = \frac{-B_2 m_1 m_3}{2c_{44}}, \varepsilon_{23} = \frac{-B_2 m_2 m_3}{2c_{44}} \quad \varepsilon_{33} = \frac{B_1 - 3c_{12}(\varepsilon_{11}^{res} + \varepsilon_{22}^{res}) - 3B_1 m_3^2}{3c_{11}}. \tag{7}$$

The time-varying magnetization associated with the Kittel mode magnon is written as $\mathbf{m} = \mathbf{m}^0 + \Delta\mathbf{m}(\mathbf{k} = 0, t) = (m_1^0 + \Delta m_1^0 e^{-i\omega t}, m_2^0 + \Delta m_2^0 e^{-i\omega t}, m_3^0 + \Delta m_3^0 e^{-i\omega t})$, where $\mathbf{m}^0 = (1,0,0)$ is the initial equilibrium magnetization and $\omega$ is the angular frequency.

By substituting the Eqs. (2-3) as well as the expressions of $\mathbf{m}$ into the LLG equation (Eq. (1)), setting the Gilbert damping coefficient $\alpha$ to zero, and dropping higher order terms, the LLG equation can be converted into a set of linear equations. By letting the determinant of the coefficient matrix of these linearized equations to be zero (also see details in Ref. 23), one can derive the angular resonant frequency of the Kittel magnon (see Supplementary Information), given by,

$$\omega = \frac{\gamma\sqrt{2K_1 + H_1^{bias} M_s \mu_0}}{\sqrt{3} M_s \mu_0} \sqrt{\frac{2B_1^2}{c_{11}} - \frac{3B_2^2}{c_{44}} + 6K_1 + 3M_s \mu_0 (H_1^{bias} + M_s) - \frac{6B_1}{c_{11}}(c_{11} + 2c_{12})\varepsilon_{res}} \tag{8}$$

where $\varepsilon_{res} = \varepsilon_{11}^{res} = \varepsilon_{22}^{res}$, i.e., the in-plane biaxial residual strains are isotropic. Equation (8) is fitted to all measured FMR data allowing for the extraction of both $K_1$ and $B_1$ (hence the $\lambda_{100}$). The $B_2$ is set to a constant because the contributions from $B_2$ to the magnetoelastic coupling is limited since $m_2, m_3 \to 0$.

Figure 2 shows the obtained FMR curves. The value of magnetization saturation was set equal to values measured by vibrating sample magnetometry (VSM) to allow for a greater accuracy in the determination of $K_1$ and $\lambda_{100}$ from the fit. The frequency dependent FMR data fits well to the Kittel equation (8).

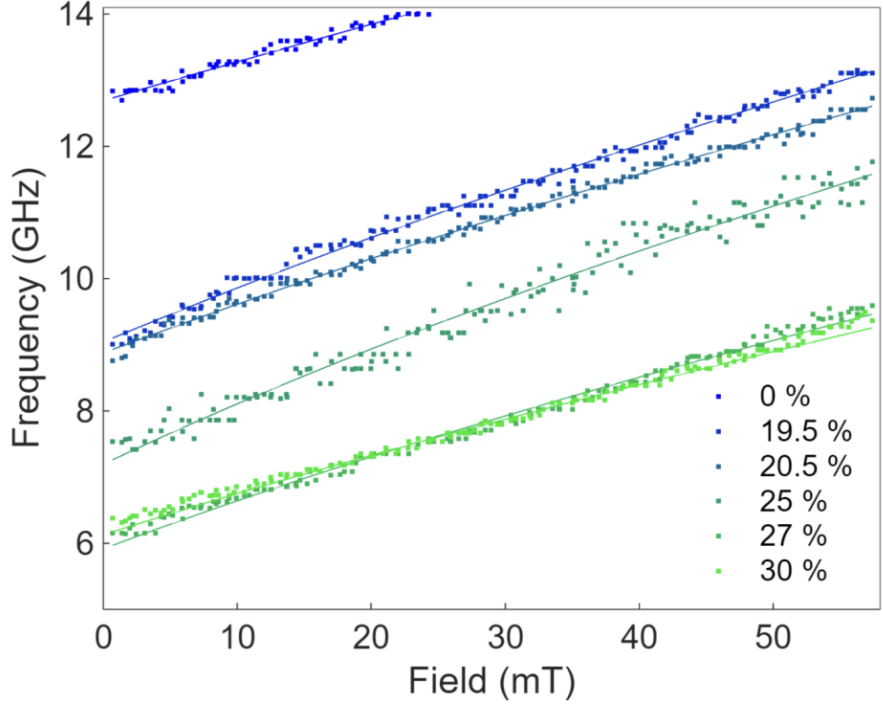

Figure 2: Measured FMR from the A2 FeGa films of varied Ga atomic percent (points) and the resulting least-squares fits with the Kittel equation (8) with magnetic anisotropy (solid lines). The Kittel equation (8) fits the data well over the field and frequency ranges considered here.

The fitted values of $K_1$ and $B_1$ as well as the values of $B_2$, $c_{11}$, $c_{12}$, $c_{44}$, and $\varepsilon_{res}$ are all listed in the supplementary material. The measured values of $M_s$ and $K_1$ from magnetometry (detailed in the supplementary material) can be seen in Figure 3 alongside the fitted values of $K_1$ and $\lambda_{100}$ from FMR. $M_s$ values systematically decrease with increasing Ga concentration as expected for non-magnetic alloy dilution. The $K_1$ values from the magnetometry and the Kittel fit agree at lower Ga% but disagree at higher values where the measured values are close to zero and the fitted values increase dramatically. The decreasing magnetic anisotropy constant with increasing Ga% from magnetometry suggests that domain wall pinning in the material is reducing. Additionally, $\lambda_{100}$ increases with increasing Ga%, which agrees with prior values from literature[1,3] and shows the high magnetostrictive coefficient for which these films are known.

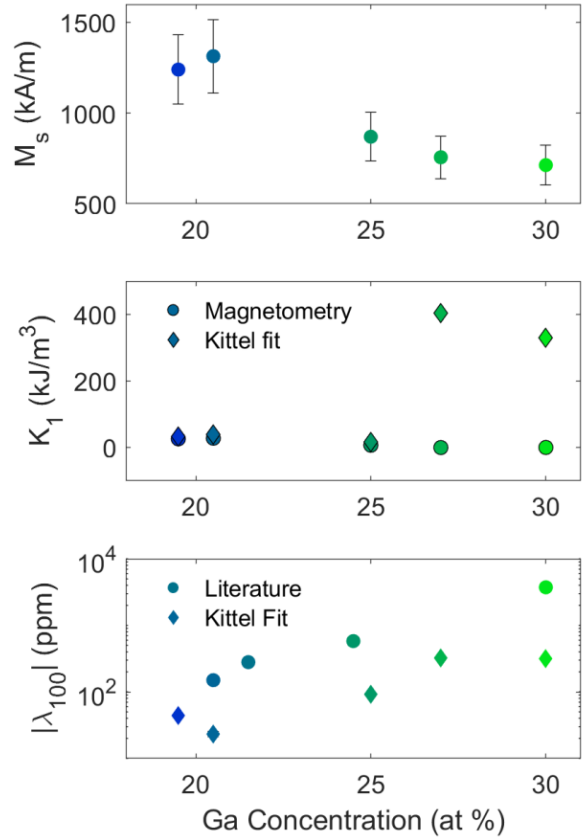

Figure 3: (a) Saturation magnetization as a function of atomic % Ga measured by VSM with standard deviation calculated as a function of the error from different measurements. (b) Anisotropy constants from VSM and Kittel equation (8) fits as a function of atomic % Ga. Magnetometry errors were determined by the standard deviation in the measurement and the standard deviation of the Kittel fit to the experimental data, both of which are too small to be seen in the figure. (c) The magnetostriction coefficient ($\lambda_{100}$) as a function of atomic % Ga from the Kittel fits shown in Figure 2 and from literature[1,3]. The error bars (which are too small to be seen in this figure) show the fit error.

In order to derive the Gilbert damping parameter for these samples, the linewidth of the FMR curves is measured and fitted to the below equation[27].

$$\Delta H = \left(\frac{2\pi}{\gamma}\right) 2\alpha\omega + \Delta H_0 \qquad (9)$$

The values of α that come from the fit are shown in Figure 4. The value for α that was measured for the x = 0 at% sample was not included due to the limits of our measuring instruments. Most of the FMR signal for the pure Fe sample extended outside the frequency range of our VNA (1-14 GHz) making it impossible to accurately measure the FMR linewidth. The standard deviation values shown were calculated from the VNA measurement error, with the largest standard deviation being around 18% of the measured value for x = 25 at%, which resulted from its being the sample with the noisiest FMR measurements.

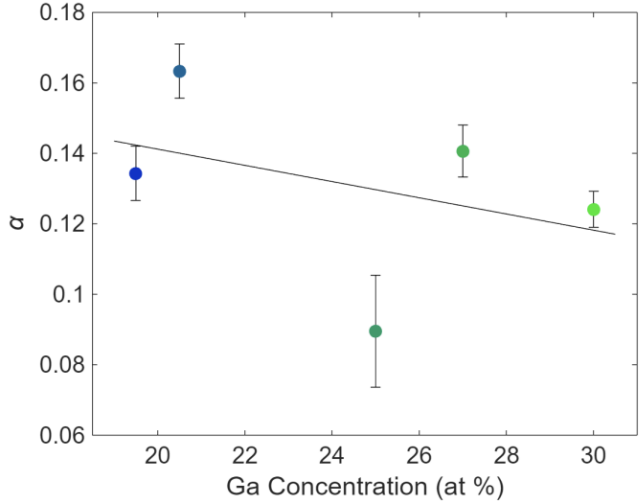

Figure 4: Gilbert damping parameter ($\alpha$) as a function of at % Ga for x = 19.5 to 30 at% with a linear fit to the data. The value of $\alpha$ decreases by 0.002 for every increase in Ga at%. The standard deviation was calculated as a function of the standard deviation from the VNA measurement error.

The correlation between Ga at% and α displays an overall linear trend in this composition range, with α decreasing by 0.002 for every increase in Ga at% as seen in the linear fit in Figure 4. As the magnetostriction increases in these samples, the Gilbert damping decreases, which makes these epitaxially-stabilized films excellent candidates for low-damping composite magnetoelectric devices.

We now attempt to provide a more quantitative analysis of the total effective Gilbert damping, which, heuristically, can be written as $\alpha_{tot} = \alpha_{ME} + \alpha_{NMS} + \alpha_{2MS}$ based on the different mechanisms of magnon scattering discussed in Ref. 28. Here, $\alpha_{ME}$ is the intrinsic contribution from the magnon-phonon scattering, $\alpha_{NMS}$ is the contribution from other intrinsic dissipation channels apart from the magnon-phonon scattering, such as the magnon-electron scattering; $\alpha_{2MS}$ is the contribution of the two-magnon scattering which arises from the presence of magnetic/structural inhomogeneity that scatters the **k** = 0 Kittel magnon into degenerate **k**≠0 magnons[29]. In general, larger magnetoelastic coupling coefficients lead to larger magnetic energy dissipation during the excitation and relaxation process (see exact form of the dissipated energy in Ref. 24), thus leads to a higher $\alpha_{ME}$. An expression of $\alpha_{ME}$ is provide in Ref. 20, reproduced below in SI units,

$$\alpha_{ME} = \frac{36\rho\gamma}{\mu_0 M_s \tau}\left(\frac{\lambda_{100}^2}{q_L^2} + \frac{\lambda_{111}^2}{q_T^2}\right), \tag{10}$$

where $\rho$ is the mass density, $\tau$ is the acoustic absorption coefficient, $q_T \approx v_T M_s / 2\gamma A_{ex}$ and $q_L \approx q_T v_T / v_L$ are the longitudinal and transverse acoustic propagation constants, respectively. $A_{ex}$ is the exchange stiffness constant, and $v_L$ and $v_T$ are the longitudinal and transverse acoustic velocities, respectively. Plugging in these values for FeGa (see Supplementary Material), $\alpha_{ME}$ for Ga at 19.5%, 20.5%, 25%, 27%, and 30% is calculated to be $1.49\times10^{-3}$, $5.42\times10^{-4}$, 0.0145, 0.225, and 0.238, respectively. Overall, $\alpha_{ME}$ increases as the Ga composition increases, and the fact that the ratio of $\alpha_{ME}/\alpha_{tot}$ exceeds 1 at high Ga composition (27% and 30%) may indicate the limited scope of the application of Eq. (10). Despite this, Equation (10) is consistent with the physical intuition that Gilbert damping coefficient should increase with $\lambda_{100}$ (more precisely, |$B_1$|). Because $\lambda_{100}$ generally increases with Ga composition [Fig. 3(c)], the observed reduced damping at high Ga composition cannot be explained from the magnon-phonon scattering alone. Rather, the other magnon scattering mechanisms (e.g., $\alpha_{NMS}$ and/or $\alpha_{2MS}$ mentioned above) are likely to play a more dominant role, especially at high Ga compositions.

In summary, we report on flip-chip FMR measurements of epitaxially-stabilized Fe$_{1-x}$Ga$_x$ thin films at room temperature with compositions spanning x = 19.5 to 30 at%. The Gilbert damping parameter decreases with increasing Ga fraction. At x = 30 at% the Gilbert damping parameter was found to be ~0.13. Due to the decrease in magnetic anisotropy, decrease in damping parameter, and increase in magnetostriction with increasing Ga fraction, the epitaxially-stabilized FeGa thin films, especially at 30% Ga, are excellent candidates for magnetoelectric composite materials.

See the Supplementary material for information about the magnetometry and magnetic anisotropy measurements and the calculation of the Kittel fit.

**Acknowledgements**

This work was made possible by the National Science Foundation SWIFT program (EECS 2229440) and the MSTAR program supported by the State of Michigan. J.-M.H. acknowledges the support from the U.S. National Science Foundation (NSF) under the Award No. DMR-2509513 (J.-M.H.). J.W. and J.-M.H. also acknowledge the partial support of this research by NSF through the University of Wisconsin Materials Research Science and Engineering Center (DMR-2309000).

**Author Declarations**

The authors have no conflicts to disclose. J.H. proposed the study which was supervised by J.H., C.A., F.C., and J.-M.H. R.L. performed the FMR measurement and analyzed the data. S.S. performed magnetometry, XRD and XRR and grew the films. J.W. derived the Kittel model and calculated the fitting values under the supervision of J.-M.H.. R.L. and J.W. wrote the manuscript. All authors commented on the manuscript and contributed to its final version.

**Data Availability Statement**

The data that support the findings of this study are available from the corresponding author upon reasonable request.

A meme summarizing part of this work can be found on https://ferroelectronicslab.com/memes/

## Magnetometry

Magnetometry data was collected along the [110] and [100] FeGa crystallographic orientations at room temperature using a vibrating sample magnetometer (VSM). The anisotropy energy $k_1$ was calculated from the difference in the integration of the moment versus field curves $A_{ijk} = \int_0^{H(M_s)} MdH$ as described by $A_{110} - A_{100} = K_1/4$. The $k_1$ values measured from the magnetometry differ from the values calculated from the Kittel fitting of the FMR curves as seen in Figure 4.

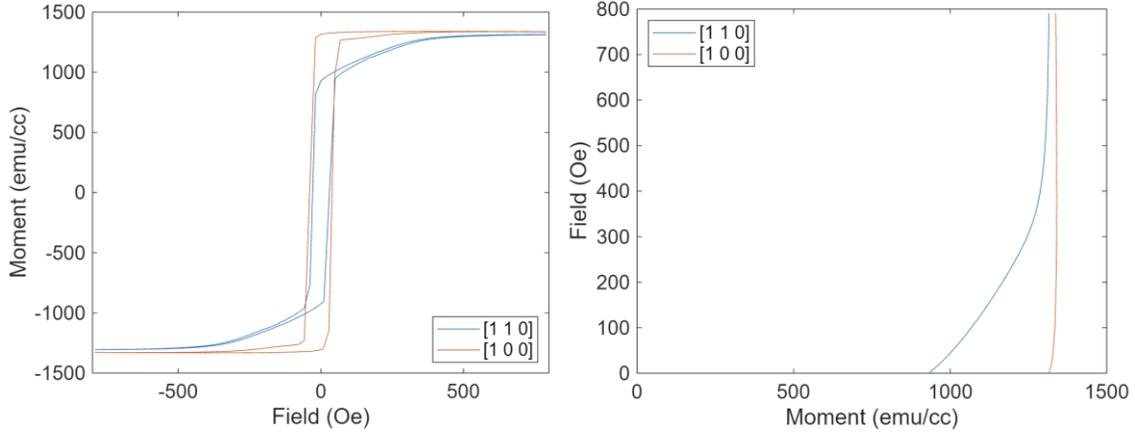

Figure 1S: (left) Magnetometry data for the x = 20.5 Ga at % sample. (right) Magnetometry data in a different orientation showing the two curves used to calculate the magnetic anisotropy.

## Kittel Fitting

By substituting **m** and $\varepsilon_{ij}$ into the LLG equation (Eq. 1), neglecting the damping α, and dropping the higher-order terms, the LLG equation is linearized to the form,

$$\begin{pmatrix} i\omega + A_{11} & A_{12} & A_{12} \\ A_{21} & i\omega + A_{22} & A_{23} \\ A_{31} & A_{32} & i\omega + A_{33} \end{pmatrix} \begin{pmatrix} \Delta m_x^0 e^{-i\omega t} \\ \Delta m_y^0 e^{-i\omega t} \\ \Delta m_z^0 e^{-i\omega t} \end{pmatrix} = 0, \qquad (S1)$$

where the explicit forms of $A_{ij}$ are not shown here for their complexity. The FMR resonance condition can be determined by setting the determinant of the coefficient matrix in Equation S2 to zero,

$$\det \begin{pmatrix} i\omega + A_{11} & A_{12} & A_{12} \\ A_{21} & i\omega + A_{22} & A_{23} \\ A_{31} & A_{32} & i\omega + A_{33} \end{pmatrix} = 0 \qquad (S2)$$

In our case, we consider that the bias field $\boldsymbol{H^{bias}}$ and the internal magnetization $m^0$ are both along '1' direction, and that the in-plane mismatch strain satisfies $\varepsilon_{11}^{res} = \varepsilon_{22}^{res} = \varepsilon_{res}$. The non-zero terms of $A_{ij}$ reduce to:

$$A_{23} = -(H_1^{bias} + M_s)\gamma - \frac{2B_1^2\gamma}{3c_{11}M_s\mu_0} + \frac{B_2^2\gamma}{c_{44}M_s\mu_0} - \frac{2K_1\gamma}{M_s\mu_0} + \left(1 + \frac{2c_{12}}{c_{11}}\right)\frac{2B_1\gamma}{M_s\mu_0}\varepsilon^{res},$$

$$A_{32} = H_1^{bias}\gamma + \frac{2K_1}{M_s\mu_0}.$$

The FMR resonance condition is:

$$\omega = \frac{\gamma\sqrt{2K_1+H_1^{bias}M_s\mu_0}}{\sqrt{3}M_s\mu_0}\sqrt{\frac{2B_1^2}{c_{11}} - \frac{3B_2^2}{c_{44}} + 6K_1 + 3M_s\mu_0(H_1^{bias} + M_s) - \frac{6B_1}{c_{11}}(c_{11} + 2c_{12})\varepsilon_{res}} \quad (S3)$$

This relationship is fitted to all measured FMR data allowing for the extraction of both $k_1$ and $\lambda_{100}$ where $B_1 = -1.5\lambda_{100}(c_{11} - c_{12})$ and $B_2 = -3\lambda_{111}c_{44}$

The parameters used for the fitting are outlined in the table below.

| Ga at% | 0% | 19.5% | 20.5% | 25% | 27% | 30% |
|---|---|---|---|---|---|---|
| a (nm) | 0.2828* | 0.2858* | 0.286* | 0.2867* | 0.287 | 0.2875* |
| c (nm) | 0.2879 | 0.2907 | 0.2914 | 0.2920 | 0.2921 | 0.2925 |
| $\varepsilon^{res}$ | -1.4704e-2 | -1.5079e-2 | -1.5098e-2 | -1.5184e-2 | -1.5222e-2 | -1.5278e-2 |
| $c_{11}$ (GPa) | 226 [3] | 198[2] | 196[2] | 184 [2] | 178 [2] | 170 [2] |
| $c_{12}$ (GPa) | 140 [3] | 140 [2] | 140 [2] | 139 [2] | 138 [2] | 137 [2] |
| $c_{44}$ (GPa) | 116 [3] | 122* | 122 [4] | 123* | 124* | 125* |
| $\gamma$ (rad*Hz*m/A) | 2.25e5 [1] | 2.25e5 [1] | 2.25e5 [1] | 2.25e5 [1] | 2.25e5 [1] | 2.25e5 [1] |
| $M_s$ (kA/m) | 1380.9 | 1239.9 | 1313.6 | 869.6 | 756.3 | 713.2 |
| $\lambda_{111}$ | -22.7e-6 [5] | 25.5e-6* | 28e-6 [4] | 39.1e-6* | 44.1e-6* | 51.5e-6* |

*Values marked with an asterisk are estimated from values for other compositions extracted for that specific composition with a linear fit.

The fitted parameters are summarized as follows:

| Ga at% | 0% | 19.5% | 20.5% | 25% | 27% | 30% |
|---|---|---|---|---|---|---|
| $K_1$ (kJ/m$^3$) | 78.1 | 34.0 | 39.8 | 16.6 | 404 | 329 |
| $\lambda_{100}$ (ppm) | 15.4 | -43.7 | 23.3 | -91.2 | 321 | 319 |
| $B_1$ (MJ/m$^3$) | -1.99 | 3.80 | -1.96 | 6.16 | -19.3 | -15.8 |

The parameters used for estimation of Gilbert damping arising from magnetoelastic coupling [6] are listed below. $\Gamma = 2.25\times10^5$ rad Hz m/A is assumed to be independent of composition.

| Ga% | 0% | 19.5% | 20.5% | 25% | 27% | 30% |
|---|---|---|---|---|---|---|
| $\rho$ (kg/m$^3$) | 7874 | 7804 | 7800 | 7784 | 7777 | 7766 |
| $c_{11}$(GPa) | 226 [3] | 198 [2] | 196 [2] | 184 [2] | 178 [2] | 170 [2] |
| $c_{12}$(GPa) | 140 [3] | 140 [2] | 140 [2] | 139 [2] | 138 [2] | 137 [2] |
| $c_{44}$(GPa) | 116 [3] | 122* | 122 [4] | 123* | 124* | 125* |
| $v_L = \sqrt{c_{11}/\rho}$ (m/s) | 5357 | 5037 | 5013 | 4862 | 4784 | 4679 |
| $v_T = \sqrt{c_{44}/\rho}$ (m/s) | 3838 | 3954 | 3955 | 3975 | 3993 | 4012 |
| $q_T = \frac{v_T M_s}{2\gamma A_{ex}}$ (m$^{-1}$) | 7.048 | 7.543 | 8.060 | 5.567 | 4.946 | 4.811 |

| $q_L = q_T v_T / v_L$ (m⁻¹) | 5.050 | 5.921 | 6.359 | 4.551 | 4.128 | 4.125 |
| --- | --- | --- | --- | --- | --- | --- |
| $\tau$ (10⁻¹³ s) [6] | 1.8 | 1.8 | 1.8 | 1.8 | 1.8 | 1.8 |
| $\alpha_{ME} = \frac{36\rho\gamma}{\mu_0 M_s \tau}\left(\frac{\lambda_{100}^2}{q_L^2} + \frac{\lambda_{111}^2}{q_T^2}\right)$ (SI) | 5.10e-4 | 1.49e-3 | 5.42e-4 | 0.0145 | 0.225 | 0.238 |

*Values marked with an asterisk are estimated from values for other compositions extracted for that specific composition with a linear fit.